\documentstyle[sprocl,epsfig]{article}

\def\Journal#1#2#3#4{{#1}~{\bf #2}, #3 (#4)}

\def\NPA{{\em Nucl. Phys.} A}
\def\NPB{{\em Nucl. Phys.} B}
\def\PLB{{\em Phys. Lett.} B}

\def\PRD{{\em Phys. Rev.}~D}
\def\ZPC{{\em Z. Phys.} C}
\def\ZPA{{\em Z. Phys.} A}
\def\PREP{{\em Phys. Rep.} }
\def\JPG{{\em J. Phys.} G}

\def\be{\begin{equation}}
\def\ee{\end{equation}}
\def\bea{\begin{eqnarray}}
\def\eea{\end{eqnarray}}

\def\ageq{\vcenter{\vbox{\hbox{$\buildrel > \over \sim$}}}}
\def\preprint#1{\vskip -150pt \noindent\hfill\hbox{#1}\vskip 140pt}

\begin{document}

\title{EXCLUSIVE PHOTO- AND ELECTROPRODUCTION \\ OF MESONS IN THE GEV REGION
\footnote{Talk given by W. Schweiger
at the workshop \lq\lq Diquarks 3\rq\rq,
Torino, Italy, Oct. 1996.
}
}
\author{ G. FOLBERTH, R. ROSSMANN, AND W. SCHWEIGER}

\address{Institute of Theoretical Physics, University of Graz,
Universit\"atsplatz 5,
\\ A-8010 Graz, Austria}


\maketitle
\preprint{UNIGRAZ-UTP 12-02-97}

\abstracts{
We consider the reactions $\gamma\, {\rm p}\, \rightarrow \, {\rm M}\, {\rm
B}$, with ${\rm M}\, {\rm B}$ being either ${\rm K^{+}}\, \Lambda$,  ${\rm
K}^+\, \Sigma^0$, or $\pi^+\, {\rm n}$, within a diquark model which is based
on
perturbative QCD. The model parameters and the quark-diquark distribution
amplitudes of
the baryons are taken from previous investigations of electromagnetic baryon
form
factors and Compton-scattering off protons. Reasonable agreement with the few
existing
photoproduction data at large momentum transfer is found for meson distribution
amplitudes compatible with the asymptotic one ($\propto x (1-x)$). We present
also
first results for hard electroproduction of the ${\rm K^{+}}\, \Lambda$ final
state.
Our predictions exhibit some characteristic features which could easily be
tested in
forthcoming electroproduction experiments.
}

\section{Introduction}
Our investigation of exclusive photo- and electroproduction of mesons is part
of a
systematic study of hard exclusive reactions \cite{Ja93,Kro93,Kro96}  within a
model
which is based on perturbative QCD, in  which baryons, however, are treated as
quark-diquark systems. This model has already been applied to baryon form
factors in the space- \cite{Ja93} and time-like region \cite{Kro93}, real and
virtual
Compton scattering \cite{Kro96}, two-photon annihilation into proton-antiproton
\cite{Kro93} and the charmonium decay $\eta_{{\rm c}} \rightarrow {\rm p}\,
\bar{{\rm
p}}$ \cite{Kro93}. A consistent description of these reactions has been
achieved in the
sense that the corresponding large momentum-transfer data ($p_\perp^2 \ageq
3$~GeV$^2$)
are reproduced with the same set of model parameters. Further applications
of this model include three-body J/$\Psi$ decays \cite{Kada95} and the
calculation of
Landshoff contributions in elastic proton-proton scattering \cite{Ja94}.
Like the usual hard-scattering approach (HSA) \cite{BL89} the diquark-model
relies on
factorization of short- and long-distance dynamics; a hadronic amplitude is
expressed
as a convolution of a hard-scattering amplitude $\widehat{T}$, calculable
within
perturbative QCD, with distribution amplitudes (DAs) $\phi^{\rm H}$ which
contain the
(non-perturbative) bound-state dynamics of the hadronic constituents. The
introduction
of diquarks does not only simplify computations, it is rather motivated by the
requirement to extend the HSA from (asymptotically) large down to intermediate
momentum transfers ($p_\perp^2 \ageq 3$~GeV$^2$). This is the momentum-transfer
region
where some experimental data already exist, but where still persisting
non-perturbative effects, in particular strong correlations in baryon wave
functions,
prevent the pure quark HSA to become fully operational. Diquarks may be
considered as
an effective way to cope with such effects. Photoproduction of mesons is one
class of exclusive reactions where the pure-quark HSA obviously fails in
reproducing the hard-scattering data \cite{FHZ91} and where a modification of
the
perturbative QCD approach seems to be necessary in the few-GeV region.

The following section starts with an outline of the hard-scattering approach
with diquarks and introduces the hadron DAs to be used in the sequel.
The diquark model predictions for the various reaction channels are presented
in Sec.~\ref{sec:photoprod} along with some general considerations on photo-
and
electroproduction reactions. Our
conclusions are finally given in Sec.~\ref{sec:conclusion}.

\section{Hard Scattering with Diquarks}
\label{sec:diquark}
\begin{figure}
\begin{center}
\epsfig{file=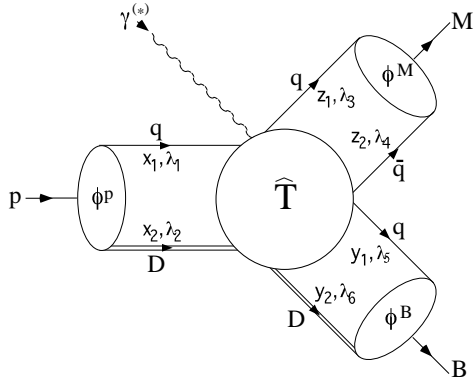, height=5cm}
\end{center}
\caption{Graphical representation of the  hard-scattering formula,
Eq.~(\ref{convol}),
for $\gamma \, {\rm p} \rightarrow {\rm M} \, {\rm B}$. $x_i$, $y_i$, and $z_i$
denote
longitudinal momentum fractions of the constituents, the $\lambda_i$'s their
helicities. }
\label{kinem}
\end{figure}

\begin{figure}
\begin{center}
\epsfig{file=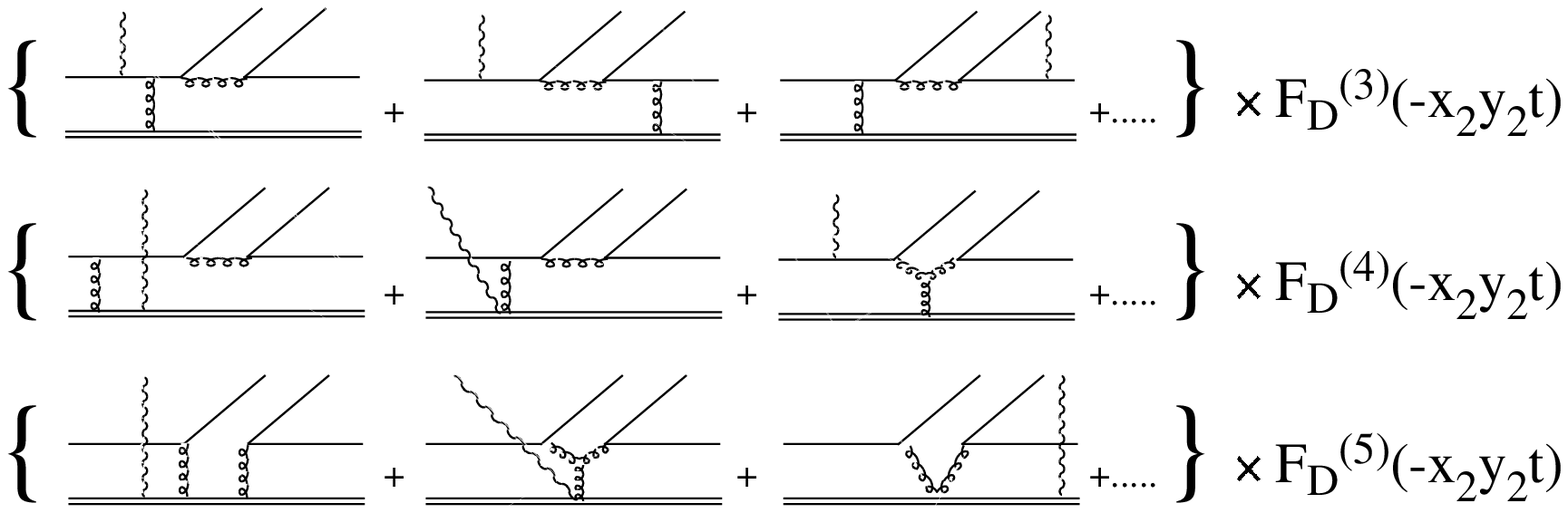, width=4.7in}
\end{center}
\caption{A few representative examples for three-, four-, and five-point
contributions to $\gamma \, {\rm p} \rightarrow {\rm M} \, {\rm B}$. As
indicated, diagrams are first calculated in Born order for point-like diquarks
and
afterwards multiplied with appropriate vertex functions (diquark form
factors).}
\label{feyn}
\end{figure}

Within the hard-scattering approach a helicity amplitude $M_{\{\lambda\}}$ for
the reaction $\gamma^{(\ast)} \, {\rm p} \rightarrow {\rm M} \, {\rm B}$ is (to
leading order in $1/p_\perp$) given by the convolution integral~\cite{BL89}
\begin{equation}
M_{\{\lambda\}}(\hat{s},\hat{t}) \! = \! \int_0^1 \! \! dx_1 dy_1 dz_1
{\phi^{\rm M}}^{\dagger}(z_1,\tilde{p}_{\! \perp})
{\phi^{\rm B}}^{\dagger}(y_1,\tilde{p}_{\! \perp})
\widehat{T}_{\{\lambda\}}(x_1,y_1,z_1;\hat{s},\hat{t})
\phi^{\rm p}(x_1,\tilde{p}_{\! \perp}) \, . \label{convol}
\end{equation}
The distribution amplitudes $\phi^{{\rm H}}$ are probability amplitudes for
finding the valence Fock state in the hadron H with the constituents carrying
certain
fractions of the momentum of their parent hadron and being  collinear up to a
factorization scale $\tilde{p}_{\perp}$. In our model the valence Fock state of
an ordinary
baryon is assumed to consist of a quark (q) and a diquark (D). We fix our
notation in
such a way that the momentum fraction appearing in the argument of $\phi^{{\rm
H}}$ is
carried by the quark -- with the momentum fraction of the other constituent
(either
diquark or antiquark) it sums up to 1 (cf. Fig.~\ref{kinem}). For our actual
calculations the (logarithmic) $\tilde{p}_{\perp}$ dependence of the DAs is
neglected
throughout since it is of minor importance in the restricted energy range we
are
interested in. The hard scattering amplitude $\widehat{T}_{\{\lambda\}}$ is
calculated
perturbatively in collinear approximation and consists in our particular case
of all
possible tree diagrams contributing to the elementary scattering process
$\gamma {\rm q} {\rm D} \rightarrow {\rm q} \bar{{\rm q}} {\rm q} {\rm D}$. A
few examples of such diagrams are depicted in Fig.~\ref{feyn}. The subscript
${\{\lambda\}}$ represents the set of possible photon, proton and $\Lambda$
helicities.
The Mandelstam variables $s$ and $t$ are written with a hat to indicate that
they are
defined for vanishing hadron masses. The calculation of the hard scattering
amplitude
involves an expansion in powers of $(m_{\rm H}/\hat{s})$ which is cut off after
${\cal
O}(m_{\rm H}/\hat{s})$. Hadron masses, however, are fully taken into account in
flux
and phase-space factors.

The model, as applied in Refs.~1-3, comprises
scalar (S) as well as axial-vector (V) diquarks. V diquarks are important, if
one
wants to describe spin observables which require the flip of baryonic
helicities.
The Feynman rules for electromagnetically interacting diquarks are just those
of
standard quantum electrodynamics. Feynman rules for strongly interacting
diquarks are
obtained by replacing the electric charge $e$ by the strong coupling constant
$g_{\rm
s}$ times the Gell-Mann colour matrix $t^a$. The composite nature of diquarks
is
taken into account by multiplying each of the Feynman diagrams with diquark
form
factors $F_{\rm D}^{(n+2)}(Q^2)$ which depend on the kind of the diquark (D =
S, V),
the number $n$ of gauge bosons coupling to the diquark, and the square of the
4-momentum $Q^2$ transferred to the diquark.  The form factors are
parameterized by
multipole functions  with the power chosen in such a way that in the limit
$p_{\perp} \rightarrow \infty$ the scaling behaviour of the pure quark HSA is
recovered.

Having outlined how diquarks are treated perturbatively, a few words about the
choice
of the quark-diquark DAs (which incorporate the q-D bound-state dynamics) are
still in
order. In Refs.~1-3 a quark-diquark DA of the form ($c_1 = c_2 = 0$ for
S diquarks, $x$ ..... longitudinal momentum fraction carried by the quark)
\be
\phi_{{\rm D}}^{{\rm B}}(x)  \propto x (1-x)^3 (1 + c_1 x + c_2 x^2)
\exp \left[ - b^2 \left( \frac{m_{{\rm q}}^2}{x} + \frac{m_{{\rm D}}^2 }{
(1-x)} \right) \right] \, , \quad {\rm D = S, V} \, ,
\label{DAp}
\ee
in connection with an SU(6)-like spin-flavour wave function, turned out to be
quite
appropriate for octet baryons B. The origin of the DA, Eq.~(\ref{DAp}), is a
nonrelativistic harmonic-oscillator wave function~\cite{Hua89}. Therefore the
masses
appearing in the exponential have to be considered as constituent masses
(330~MeV for
light quarks, 580 MeV for light diquarks, strange quarks are 150~MeV heavier
than
light quarks). The oscillator parameter $b^2 = 0.248$ GeV$^{-2}$ is chosen in
such
a way that the full wave function gives rise to a value of $600$~MeV for the
mean intrinsic transverse momentum of a quark inside a nucleon.

For meson DAs various models can be found in the literature. In order to check
the
sensitivity of our calculation on the choice of the meson DAs we employ two
qualitatively different forms. On the one hand, the asymptotic DA
\be
\phi_{\rm asy}(x) \propto x (1-x) \, ,
\label{DAKa}
\ee
which solves the $\tilde{p}_{\perp}$ evolution equation for
$\phi(x, \tilde{p}_{\perp})$ in the limit $\tilde{p}_{\perp} \rightarrow
\infty$, and,
on the other hand, the DAs
\be
\phi_{\rm CZ}^{\pi}(x) \propto \phi_{\rm asy} (2 x -1)^2 \, , \quad
\phi_{\rm CZ}^{\rm K}(x) \propto \phi_{\rm asy} [0.08 + 0.6 (2 x -1)^2 - 0.25
(2 x
-1)^3] \, ,
\label{DAKCZ}
\ee
which have been proposed in Ref.~9 on the basis of QCD sum rules. The
\lq\lq normalization\rq\rq~of the meson DAs is determined by the experimental
decay
constants for the weak $\pi$, K $\rightarrow \, \mu \, \nu_{\mu}$ decays. The
analogous constants $f_{\rm S}$ and $f_{\rm V}$ for the q-D DAs of baryons are
free parameters of the model. They are, in principle, determined by the
probability
to find the q-D state (D = S, V) in the baryon B and by the transverse-momentum
dependence of the corresponding wave function.

For further details of the diquark model we refer to the publication of
R.~Jakob et
al.~\cite{Ja93}. The numerical values of the model parameters for the present
study are also taken from this paper.

\section{Photo- and Electroproduction Reactions}
\label{sec:photoprod}
Exclusive photoproduction of pseudoscalar mesons can, in general, be described
by four independent helicity amplitudes
$M_{\lambda_{\rm M},\lambda_{\rm B},\lambda_{\gamma},\lambda_{\rm
p}}$~\cite{Ba75}
\begin{eqnarray}
N = & M_{0,-\frac{1}{2},+1,+\frac{1}{2}} \, , \qquad S_1 = &
M_{0,-\frac{1}{2},+1,-\frac{1}{2}} \, ,
\nonumber \\
D = & M_{0,+\frac{1}{2},+1,-\frac{1}{2}} \, , \qquad S_2 = &
M_{0,+\frac{1}{2},+1,+\frac{1}{2}} \, .
\end{eqnarray}
$N$ , $S_1$, $S_2$, and $D$ represent non-flip, single-flip, and double-flip
amplitudes, respectively. Two additional amplitudes,
\be
\tilde{N} =  M_{0,+\frac{1}{2},0,+\frac{1}{2}} \, , \qquad
S_3 =  M_{0,-\frac{1}{2},0,+\frac{1}{2}} \, ,
\label{ML}
\ee
occur if electroproduction is considered. Electroproduction and photoproduction
are
related as far as in the one-photon approximation the electroproduction cross
section can be expressed in terms of four response functions $d\sigma_{\rm U}$,
$d\sigma_{\rm L}$, $d\sigma_{\rm T}$, and $d\sigma_{\rm I}$ for the production
of a
pseudoscalar meson by a virtual photon,
$\gamma^{\ast}\, {\rm p} \rightarrow {\rm M}\, {\rm B}$~\cite{DL72}. The two
helicity amplitudes in Eq.~(\ref{ML}) correspond to the longitudinal
polarization
degree of the virtual photon. The two response functions $d\sigma_{\rm U}$ and
$d\sigma_{\rm L}$ are just cross sections for (spin averaged) transversely and
longitudinally polarized photons, respectively. The other two functions are
transverse-transverse ($d\sigma_{\rm T}$) and longitudinal-transverse
($d\sigma_{\rm I}$) interference terms.
In the limit of vanishing photon-virtuality ($Q^2 \rightarrow 0$) $d\sigma_{\rm
U}$
reduces to the ordinary photoproduction cross section, the ratio
$d\sigma_{\rm T}/d\sigma_{\rm U}$ becomes the (negative) photon asymmetry
$\Sigma$~\cite{Ba75} and $d\sigma_{\rm L}$ as well as $d\sigma_{\rm I}$ become
zero.

{}From the q-D flavour functions (D = S, V)
\be
\chi_{\rm S}^{\rm p} = {\rm u} {\rm S}_{[{\rm u},{\rm d}]} \, , \qquad\qquad
\chi_{\rm V}^{\rm p} = \phantom{-} [{\rm u} {\rm V}_{\{{\rm u},{\rm d}\}}
-\sqrt{2} {\rm d} {\rm V}_{\{{\rm u},{\rm u}\}}] / \sqrt{3}
\, ,
\label{flavp} \ee
\be
\chi_{\rm S}^{\rm n} = {\rm d} {\rm S}_{[{\rm u},{\rm d}]} \, , \qquad\qquad
\chi_{\rm V}^{\rm n} = - [{\rm d} {\rm V}_{\{{\rm u},{\rm d}\}} -\sqrt{2} {\rm
u} {\rm V}_{\{{\rm d},{\rm d}\}}] / \sqrt{3} \, ,
\label{flavn} \ee
\be
\chi_{\rm S}^{\Sigma^0} = [ {\rm d} {\rm S}_{[{\rm u},{\rm s}]} + {\rm u} {\rm
S}_{[{\rm d},{\rm s}]} ] / \sqrt{2} \, , \quad
\chi_{\rm V}^{\Sigma^0} = [ 2 {\rm s} {\rm V}_{\{{\rm u},{\rm d}\}}- {\rm d}
{\rm V}_{\{{\rm u},{\rm s}\}} - {\rm u} {\rm V}_{\{{\rm d},{\rm s}\}}] /
\sqrt{6} \, ,
\label{flavs0} \ee
\be
\chi_{\rm S}^{\Lambda} = [ {\rm u} {\rm S}_{[{\rm d},{\rm s}]} - {\rm d} {\rm
S}_{[{\rm u},{\rm s}]} - 2 {\rm s} {\rm S}_{[{\rm u},{\rm d}]}] / \sqrt{6}  \,
,
\quad
\chi_{\rm V}^{\Lambda} =  [{\rm u} {\rm V}_{\{{\rm d},{\rm s}\}}- {\rm d} {\rm
V}_{\{{\rm u},{\rm s}\}}] / \sqrt{2} \, .
\label{flavl}
\ee
for the baryons one infers already that the three production channels we are
interested in differ qualitatively in the sense that ${\rm K}^+ \, \Lambda$ is
solely
produced via the S$_{\rm [u,d]}$ diquark, ${\rm K}^+ \, \Sigma^0$ via the
V$_{\rm
\{u,d\}}$ diquark, and $\pi^+ \, {\rm n}$ via S and V diquarks. An important
consequence of
this observation is that helicity amplitudes and hence spin observables which
require
the flip of the baryonic helicity are predicted to vanish for the \mbox{${\rm
K}^+$-$\Lambda$} final state (e.g., the polarization $P$ of the outgoing
$\Lambda$).
For the other two channels helicity flips may, of course, take place by means
of the V
diquark.

One of the qualitative features of the HSA is the fixed-angle scaling behaviour
of cross sections. The pure quark HSA implies an $s^{-7}$ decay of the
photoproduction cross section. This scaling behaviour is, of course, recovered
within
the diquark model in the limit $s \rightarrow \infty$.
However, at finite $s$, where the diquark form factors become operational
and diquarks appear as nearly elementary particles, the $s^{-7}$
power-law is modified. Additional deviations from the $s^{-7}$ decay of the
cross
section are due to logarithmic corrections which have their origin in the
running coupling constant $\alpha_{\rm s}$ and eventually in the evolution
of the DAs $\phi^{\rm H}$ (neglected in our calculation).

\subsection{The ${\rm K}^+$-$\Lambda$ Channel}
\begin{figure}[t!]
\begin{center}
\epsfig{file=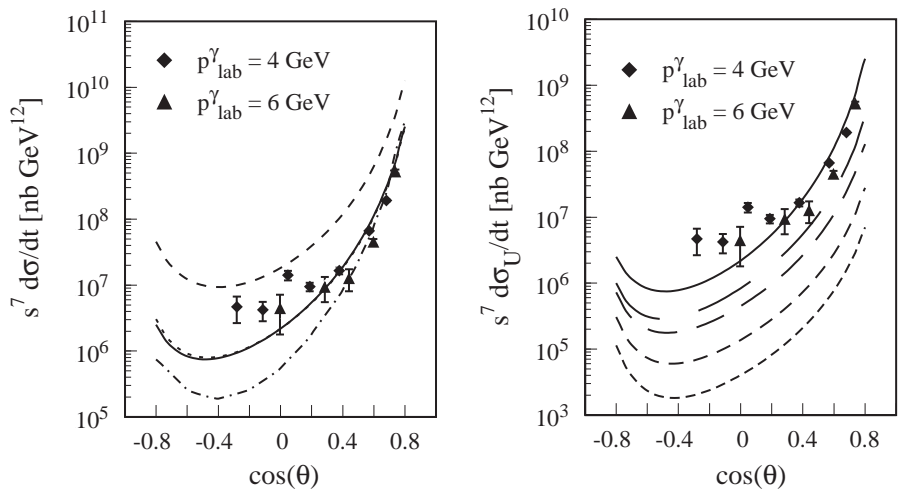, width=4.7in}
\end{center}
\caption{Diquark-model predictions for photo- (left figure) and
electroproduction
(right figure) of the K$^+$-$\Lambda$ final state. Results are shown for fixed
$p^{\gamma}_{\rm lab} = 6$~GeV. The experimental points are photoproduction
data taken from Anderson et al.\ \protect \cite{An76}.
{\em Left figure}: differential cross section for $\gamma \, {\rm p}
\rightarrow
{\rm K}^{+} \, \Lambda$ scaled by $s^7$ vs $\cos(\theta_{{\rm cm}})$; solid
(dashed) line: diquark-model result for p and $\Lambda$ DAs chosen
according to Eq.\ (\ref{DAp}), K$^{+}$ DA according to Eq.\ (\ref{DAKa})
(Eq.~(\ref{DAKCZ})); dotted line: same as full line, but only three-point
contributions taken into account;
dash-dotted line:
quark-model result \protect \cite{FHZ91} for the asymmetric p and $\Lambda$
DAs of Ref.~13 and the two-humped K$^{+}$ DA of Eq.~(\ref{DAKCZ}).
{\em Right figure}: the spin-averaged cross section $d\sigma_{\rm U}/dt$
(scaled by
$s^7$) vs $\cos(\theta_{{\rm cm}})$ for transversely polarized photons with
virtualities $Q^2 =$ 0 (photoproduction limit), 0.5, 1, 2, and 3 GeV$^2$;
dashes become  shorter with increasing $Q^2$. Proton and $\Lambda$ DAs are
chosen according to Eq.\ (\ref{DAp}), the K$^{+}$ DA according to Eq.\
(\ref{DAKa}).
}
\label{KL}
\end{figure}

Figure~\ref{KL} (left) shows the diquark-model predictions for the
(scaled) photoproduction cross section $s^7 d\sigma/dt$ along with the few
existing
photoproduction data at large-momentum transfer \cite{An76} and the outcome of
the
pure quark HSA \cite{FHZ91} (dash-dotted curve). Whereas the DAs of
proton and $\Lambda$ have been kept fixed according to Eq.~(\ref{DAp}) we have
varied
the K$^+$ DA. The solid and the dashed line represent results for the
asymptotic
(Eq.~(\ref{DAKa})) and the two-humped (Eq.~(\ref{DAKCZ})) K$^+$ DA,
respectively,
evaluated at $p_{{\rm lab}}^{\gamma} = 6$ GeV. The better
performance of the asymptotic DA and the overshooting of the asymmetric DA is
in line with the conclusion drawn from the investigation of the pion-photon
transition
form factor \cite{JKR96} where, for the case of the pion, $\phi^{\pi}_{\rm CZ}$
leads also to an overshooting the data. Our findings have to be contrasted with
those
obtained within the pure quark-model calculation of photoproduction
\cite{FHZ91},
where the asymptotic forms for both, baryon and meson DAs, give systematically
larger
results than the combination of very asymmetric DAs. However, the numerics of
Ref.~7 must be taken with some provisio. For Compton scattering off nucleons
it has been demonstrated \cite{KN91} that the very
crude treatment of propagator singularities adopted in Ref.~7,
namely keeping $i \epsilon$ small but finite, may lead to deviations from the
correct
result which are as large as one order of magnitude. At this point we want to
emphasize that we have paid special attention to a correct and numerically
robust
treatment of the propagator singularities which occur in the convolution
integral,
Eq.~(\ref{convol}). By carefully separating the singular contributions,
exploiting
delta functions, rewriting principal-value integrals as ordinary integrals plus
analytically solvable principle-value integrals, we are able to do all the
numerical
integrations by means of very fast and stable fixed-point Gaussian quadratures.

We have also examined the relative importance of various groups of Feynman
graphs and found the 3-point contributions to be by far the most important (cf.
Fig.~\ref{KL} (left)). 4- and 5-point contributions amount to $\approx 5\%$ at
$\theta_{\rm cm} = 90^{\circ}$ and $E_{\rm lab}^{\gamma} = 6$ GeV as long as
only $d\sigma/dt$ is considered.
Spin observables, on the
other hand, are much more affected by 4- and 5-point contributions. A more
detailed
discussion of
$\gamma\, {\rm p}\, \rightarrow \, {\rm K}^+\, {\Lambda}$ (and also $\gamma\,
{\rm
p}\, \rightarrow \, {{\rm K}^{\ast}}^+\, {\Lambda}$) with a full account of
calculational techniques and analytical expressions for the photoproduction
amplitudes
can be found in Ref.~16.
%
\begin{figure}[t!]
\begin{center}
\epsfig{file=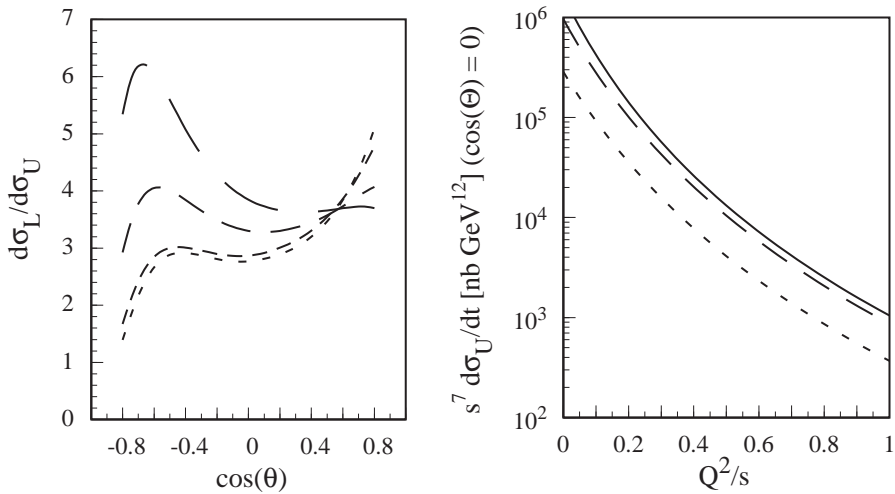, width=4.7in}
\end{center}
\caption{Diquark-model predictions for electroproduction of the K$^+$-$\Lambda$
final
state.
\mbox{
{\em Left figure}:} ratio of cross sections $d\sigma_{\rm L}/d\sigma_{\rm U}$
for
longitudinally and transversely polarized photons, respectively, vs
$\cos(\theta_{{\rm cm}})$ for fixed $p^{\gamma}_{\rm lab} = 6$~GeV; dashes
become
shorter with increasing photon virtuality $Q^2$ ($=$ 0.5, 1, 2, 3~GeV$^2$).
{\em Right figure}: $d\sigma_{\rm U}/dt$ at fixed $\theta_{\rm cm} =
90^{\circ}$ vs
$Q^2/s$ for $s =$ 10 (solid), 20 (long dashed), and 100 GeV$^2$ (short
dashed).}
\label{scal}
\end{figure}

The diquark-model predictions for a few electroproduction observables are
depicted in Fig.~\ref{KL} (right) and Fig.~\ref{scal}. In these plots we have
concentrated on the asymptotic form, Eq.~(\ref{DAKa}), of the K$^+$ DA. Like in
the
case of photoproduction the large-$s$ behaviour of the four cross section
contributions
is
$s^{-7}$ (modified by logarithmic factors) provided that $Q^2/s$ is kept fixed.
For fixed photon virtuality $Q^2$ and $s \rightarrow \infty$, however,
$d\sigma_{\rm
L}/dt$ and $d\sigma_{\rm I}/dt$ decay like $s^{-9}$ and $s^{-8}$, respectively.
For fixed  $p_{\rm lab}^{\gamma} = 6$~GeV the transverse cross-section
contribution
$d\sigma_{\rm U}/dt$ decreases with increasing $Q^2/s$ (cf.
Fig.~\ref{KL} (right)). The same holds for $d\sigma_{\rm L}/dt$ (if $Q^2/s
\ageq
0.04$). At $Q^2/s = 0.04$ $d\sigma_{\rm L}/dt$ is already larger than
$d\sigma_{\rm
U}/dt$ (cf. Fig.~\ref{scal} (left)). Since $d\sigma_{\rm L}/dt$ vanishes for
$Q^2
\rightarrow 0$ it thus has to rise very sharply at small values of $Q^2/s$.
To get a better feeling for the $Q^2$-dependence of $d\sigma_{\rm U}/dt$ we
have
plotted this quantity for fixed scattering angle ($\theta_{\rm cm} = 90
^{\circ}$) and
different values of $s$ as function of $Q^2/s$ (cf. Fig.~\ref{scal} (right)).
For exact $s^{-7}$ scaling the curves for the three different values of $s$
should
coincide. The deviation from the $s^{-7}$ scaling can mainly be ascribed to the
running coupling constant $\alpha_{\rm s}$. The dependence of $d\sigma_{\rm
U}/dt$
(and likewise $d\sigma_{\rm L}/dt$) on $Q^2/s$ can be roughly parameterized by
means
of a function $\propto (1 + c\, Q^2/s)^{-6}$ with the parameter $c$ between 2
and 2.5.
The qualitative features, like scaling behaviour, $Q^2/s$ dependence, and
dominance of
$d\sigma_{\rm L}/dt$ as compared to
$d\sigma_{\rm U}/dt$ remain, of course, unaltered if the asymptotic K$^+$ DA is
replaced by the asymmetric DA of Eq.~(\ref{DAKCZ}). The angular dependence of
cross-section ratios, like  $d\sigma_{\rm L}/d\sigma_{\rm U}$, $d\sigma_{\rm
T}/d\sigma_{\rm U}$, or $d\sigma_{\rm I}/d\sigma_{\rm U}$, however, exhibit a
marked
sensitivity on the choice of the K$^+$ DA.
First experimental constraints on hard exclusive electroproduction
are to be expected from CEBAF and later on at even higher momentum transfers
from
ELFE (at DESY).
\begin{figure}[t!]
\begin{center}
\epsfig{file=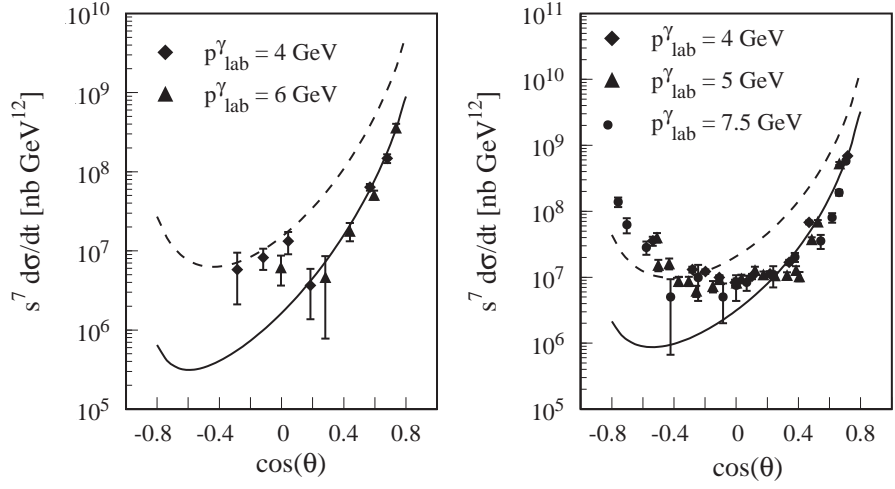, width=4.7in}
\end{center}
\vspace{-4.5 pt}
\caption{Diquark-model predictions for photoproduction of the
K$^+$-$\Sigma^0$ ({\em left figure}) and the $\pi^+$-n ({\em right figure})
final
states; solid (dashed) line: diquark-model results for baryon DAs chosen
according to Eq.\ (\ref{DAp}), K and $\pi$ DAs according to Eq.\ (\ref{DAKa})
(Eq.~(\ref{DAKCZ})). Results are shown for fixed $p^{\gamma}_{\rm lab} =
6$~GeV.
Scalar-diquark contributions are fully taken into account, four- and five-point
contributions of vector diquarks are neglected. Data are taken from
\mbox{Anderson et
al.\ \protect \cite{An76}.} }
\label{KSupin}
\end{figure}

\samepage
\subsection{The ${\rm K}^+$-$\Sigma^0$ and the $\pi^+$-${\rm n}$ Channel}
We have already mentioned that these channels differ from K$^+$-$\Lambda$
production as
far as also V-diquarks are involved. Since the treatment of V diquarks is much
more
intricate than that of S diquarks we  have, until now, only calculated the
corresponding three-point contributions. From K$^+$-$\Lambda$ production  and
also
from other applications of the diquark model we know, however, that
spin-averaged
cross sections are mainly determined by three-point contributions.
Thus we do not expect a significant modification of our numerical results for
the
K$^+$-$\Sigma^0$ and the $\pi^+$-n photoproduction cross sections if four- and
five-point contributions of V diquarks are included.  Like in the case of
K$^+$-$\Lambda$ production we observe reasonable agreement with the
K$^+$-$\Sigma^0$ and the $\pi^+$-n cross-section data if the asymptotic DA is
taken for
the K$^+$ and the $\pi^+$, respectively (cf. Fig.~\ref{KSupin}). The two-humped
DAs,
Eq.~(\ref{DAKCZ}), seem again to be in conflict with the data. Also for these
two
channels the pure quark HSA fails in reproducing the data.  It is remarkable
that the
unpolarized differential cross sections for \mbox{${\rm K}^+$-$\Lambda$} and
\mbox{${\rm K}^+$-$\Sigma^0$} production are very similar in size, although the
corresponding production mechanisms (via S and V diquarks, respectively) are
quite
different. We expect this difference to show up more clearly in spin
observables.

\newpage
\section{Conclusions}
\label{sec:conclusion}
The predictions of the diquark model for $\gamma\, {\rm p}\,
\rightarrow$ ${\rm K}^+\, \Lambda$,  ${\rm K}^+\, \Sigma^0$, and $\pi^+\, {\rm
n}$
look rather promising if the asymptotic form ($\propto x (1-x)$) is taken for
the
meson DAs. To the best of our knowledge the diquark model is, as yet, the only
constituent scattering model which is able to account for the large-$p_{\perp}$
photoproduction data. For electroproduction of the
${\rm K}^+$-$\Lambda$ final state our results are the first perturbative QCD
predictions at all.  With respect to future experiments it
would, of course, be desirable to have more and better large momentum-transfer
data
on exclusive photo- and electroproduction. Polarization measurements of the
recoiling
particle could help to decide, whether the perturbative regime has been reached
already, or whether non-perturbative effects (different from diquarks) are
still at
work. Spin observables, in general, could be very helpful to constrain the
form of the hadron DAs.

\end{document}